\documentstyle[12pt]{article}
\begin{document}
\def\om{\omega}
\def\omt{\tilde{\omega}}
\def\ti{\tilde}
\def\o{\Omega}
\def\bchi{\bar\chi^i}

\def\ba{\bar a}
\def\w{\wedge}

\def\Tr{{\rm Tr}}
\def\ST{{\rm STr}}
\def\ss{\subset}
\def\bc{{\bf C}}
\def\br{{\bf R}}

\def\al{\alpha}
\def\la{\langle}
\def\ra{\rangle}

\def\th{\theta}
\def\lm{\lambda}

\def\d{\partial}
\def\dz{\partial_z}
\def\dbz{\partial_{\bar z}}

\def\be{\begin{equation}}
\def\ee{\end{equation}}
\def\bea{\begin{eqnarray}}
\def\eea{\end{eqnarray}}
\def\A{{\cal A}}
\def\B{{\cal B}}
\def\T{{\cal T}}
\def\bT{\bar{\cal T}}
\def\Z{{\cal Z}}
\def\si{\sigma}
\def\*{\ddagger}
\def\j{\dagger}
\def\bz{\bar{z}}
\def\e{\varepsilon}
\def\b{\beta}
\def\bb{\bar b}
\def\bk{\bar k}
\begin{titlepage}
\begin{flushright}
{}~
IHES/P/97/77\\
hep-th/9710153
\end{flushright}

\vspace{3cm}
\begin{center}
{\Large \bf Gauge theories on the noncommutative sphere}\\
[50pt]{\small
{\bf C. Klim\v{c}\'{\i}k}\\ IHES, 
91440 Bures-sur-Yvette, France\\and\\Institut of Mathematics, 
University of Marseille II\\163,
Avenue de Luminy, 13288 Marseille, France}

\vskip2pc
\noindent{\it Dedicated to my son Cedrik at the occasion of his birth.}

\vspace{1cm}
\begin{abstract}
Gauge theories 
are formulated on the noncommutative
two-sphere. These theories have only finite number of degrees  of freedom,
nevertheless they exhibit both  the gauge symmetry 
and the SU(2) "Poincar\'e" symmetry of the sphere. In particular, 
the coupling of gauge fields to chiral fermions
is naturally achieved.

\end{abstract}
\end{center}

\end{titlepage}

\section{Introduction}
There was a reappearing belief through this century that a possible
understanding of a short distance behaviour of physical theories
should stem from a theory that would incorporate a minimal
lenght. At early times of  quantum field theory,
such attitude was almost dictated by the pertinent problem of ultraviolet (UV)
divergences. With the work of Wilson, we have learned better 
 to get along with
the divergences, nevertheless it is still expected that a fundamental
theory should be in a sense finite. The divergences should appear 
 in a controlled way in effective approaches to various aspects of the theory.

String theory is often presented as a candidate for a fundamental description
of our world. Its perturbation expansion is believed to be UV finite
and the string tension sets  a scale which is interpreted  as the minimal 
lenght\footnote{However, see a recent progress in understanding 
substringy scales via $D$-branes \cite{Dou}.}.  Insights of last two years
even led  to
remarkable proposals of various matrix models \cite{BFSS,IKKT} which
should constitute the nonperturbative formulation of string theory.
We should like to stress, however, that
 the idea that theories containing large matrices naturally incorporate
the concept of minimal lenght had appeared earlier in the field theoretical
context \cite{Ho,Ma,GMa,GKP2,Ke,W,DFR}. These works
stem from the basic ideas of Connes'  noncommutative geometry \cite{Con}
and argue that
a nonperturbatively well defined path integral of various 
field theories
can be formulated if the fields are replaced by big matrices in a way which
correspond to the {\it kinematical} quantization of the spacetime on which
the fields live. Because this is the crucial point of the whole matrix
approach let me be more precise here (a   reasoning quite similar
in spirit
now underlies the matrix approach to superstring theory \cite{IKKT}):

Consider a Riemann sphere as a spacetime of an Euclidean field
theory. This is a sufficiently
general setting, since we can always effectively decompactify 
the spacetime by scaling
the round metric (or radius) of the sphere 
 and  the Minkowski dynamics can be achieved by a
 suitable modification
of the standard Osterwalder-Schrader procedure. An  invariance 
of field theories with 
respect to the  
isometry group $SO(3)$
of  the sphere then in the limit
of infinite scaling play the role of the 
Euclidean
(or Poincar\'e in the Minkowski case) invariance. Now the crucial
observation is that the spacetime
$S^2$ is naturally a symplectic manifold; the symplectic form $\omega$ 
is up to a normalization
 just
the round 
volume form on the sphere. Using the standard complex coordinate
$z$ on the Riemann sphere, we have
$$\omega=-{N\over 2\pi}{d\bz \w dz\over (1+\bz z)^2},\eqno(i)$$
with $N$ a real parameter\footnote{Note, that we have chosen a normalization
which makes the form $\omega$ purely imaginary. Under quantization, hence,
the Poisson bracket is replaced by a commutator {\it without} any 
imaginary unit factor.}
. If we consider a scalar field theory,
then the scalar field $\phi$  is  a function on the symplectic manifold or,
in other words, a  classical
observable. The action of the massless (real) 
scalar field theory on $S^2$ is given by
$$S=-i\int \omega R_i\phi R_i\phi,\eqno(ii)$$
where $R_i$  are the vector fields which generate the $SO(3)$
rotations of $S^2$ and the Einstein summation convention is understood.
 The vector fields $R_i$ are Hamiltonian; this means that there
exists three concrete observables $r_i$ such that
$$ \{r_i,\phi\}=R_i\phi.\eqno(iii)$$
Here $\{.,.\}$ is the Poisson bracket which corresponds to the symplectic
structure $\omega$. The observables $r_i\in{\br}^3$ are just 
the coordinates of the embedding of $S^2$ in $\br^3$. Thus we can rewrite
the action (ii) as 
$$S=-i\int\omega\{r_i,\phi\}\{r_i,\phi\}.\eqno(iv)$$

 Suppose we quantize the symplectic structure on $S^2$
(probably the first who has done it was Berezin \cite{Ber}). Then the
algebra of observables becomes the noncommutative algebra of all square
matrices with entries in $\bc$; the quantization of $S^2$ can be only
 performed if $N$ is an integer,
the size of the scalar field matrices $\phi$ is then $(N+1)\times (N+1)$.
This algebra of matrices defines the noncommutative (or fuzzy \cite{Ma})
sphere. The integration over the phase space
volume 
form $i\omega$ is replaced by 
taking a properly normalized trace $\Tr$ over the matrices and the 
Poisson brackets are replaced by 
 commutators (the Hamiltonians $r_i$
are also quantized, of course).

 Putting together, we can  consider along with (iv) a noncommutative 
action
$$S=-{1\over N+1}\Tr([r_i,\phi],[r_i,\phi]).\eqno(v)$$

 The action (v) has a few nonstandard properties.
 First of all, the space of all "fields"(=matrices)
is finite dimensional and the product of fields is noncommutative.
The latter property may seem awkward but in all stages of analysis we shall
never encounter a problem which this noncommutativity might create. The former
property, however, is highly desirable, since all divergences of the usual
field theories are automatically eliminated. We may interpret (v) as the
regularized version of (iv); the fact that (v) goes to (iv) in the
limit $N\to\infty$ is just the statement that classical mechanics is the
limit of the quantum one for the value of the Planck constant $1/N$ approaching
zero. Remarkably, unlike lattice regularizations, (v) preserves the
"Poincar\'e" symmetry $SO(3)$ of the spacetime. Indeed, under the variation
$\delta\phi=[r_i,\phi]$ the action (v) remains invariant.

Few more words are useful
for understanding the nature of the commutative limit $N\to\infty$ 
(see \cite{GKP2} for more details). Both classical
commutative fields and their noncommutative counterparts (matrices)
can be decomposed into spherical harmonics. This means
that the algebra of observables is  a representation of the group $SO(3)$,
which infinitesimally
acts on $\phi$ via Poisson bracket or commutators (respectively)
with the Hamiltonians $r_i$. All spherical harmonics
with the quantum number $l$ form the irreducible representation
of $SO(3)$ (this is their definition in the noncommutative case).
Recall that the Laplace operator on the sphere (the Casimir $R_iR_i$
of $SO(3)$)
has the spectrum $l(l+1)$.
 Now if a maximal $l$ of $\phi$ in (v) is much smaller than $N$, then 
 the action
 (v) differs from (iv) by  a factor which goes like $1/N$
 for $N\to\infty$. This means that, being at large scales
 (=small momenta $l$), the actions (iv) and (v) are equivalent.
 They differ only in the short distance limit. Hence we stress that 
 there is no point in saying which one is better, or which action
  is an approximation
  of the other. Just from the historical reasons people
 were using (iv) earlier than (v) (they did not know about 
 the noncommutative geometry). Both (iv) and (v) are extrapolations
 of the same long distance quantity also to the ultraviolet region. 
 The symmetry principle cannot select one of them. Therefore we
  tend to believe
 that (v) might be a better choice, because it is from the outset manifestly
 regular. Such a statement may call for a criticisms; the most obvious
 would be: what about the other fields (spinor,vector etc.); can one
 construct classical field theories on noncommutative manifolds involving
 those fields? This paper constitutes 
  a step of the program \cite{GKP2,Haw,Mad2}
  which aims to show that the answer to this question is 
  positive.

Scalar fields, being observables,
are naturally defined in the noncommutative context
just as elements of the algebra
which defines the noncommutative manifold. However,   no canonical 
procedure appears to exist of 
how to desribe, say, spinor fields on a noncommutative manifold.
The early approaches by Grosse and Madore \cite{GMa} and later by
Grosse and  Pre\v snajder \cite{GP} considered spinors on $S^2$ as 
 two component
objects where both components are elements of the noncommutative algebra
of matrices $(N+1)\times(N+1)$. Although this approach did permit to construct
field theories (in the case of \cite{GMa} also the gauge fields with an 
inevitable
addition of one propagating scalar),
the noncommutative actions obtained in this way have somewhat lost
their competitivness with respect to the classical commutative actions.
The reason was that the Dirac operator did not anticommute with the chiral
grading for finite $N$. A shift of a point of view
was presented by Grosse, Klim\v c\'\i k  and Pre\v snajder in \cite{GKP2},
where it was argued that spinors should be understood as odd parts of
scalar superfields. Then the chirality of the Dirac operator
is automatically preserved. Thus the guiding principle for understanding 
the spinors
consists in quantizing
the supersymplectic manifolds. If they are compact (like a supersphere 
in two dimension) then the quantized algebra of observables is just
an algebra of supermatrices with a finite size. The odd (off-diagonal)
part of the supermatrix then describes a spinor on the bosonic submanifold
of the whole supermanifold. In this way, the scalar superfields on 
the
supersphere were shown to describe both scalars and 
 spinors on the sphere \cite{GKP2}.
The moral of the story is that, by using supersymmetry,  the
construction of noncommutative spinors is as canonical as that of
noncommutative scalars.

The story of noncommutative
gauge theories on $S^2$, compatible with the chirality of the Dirac
operator and with the already known description of the scalars and spinors,
is so far missing.
  We are going to fill   this gap
 in this article. Our strategy
will consist
in converting the standard derivatives into covariant ones
in the noncommutative actions
like (v). We shall do it in the supersymmetric
framework which treats the scalars and spinors on the
same footing and which uses
only   Hamiltonian vector fields as the derivatives
appearing in the Lagrangian. This means that we shall covariantize these (odd) 
 derivatives by adding suitable gauge fields. Then we identify
a gauge invariant kinetic term for these gauge fields 
which, remarkably, can be
also written solely in terms of the same odd Hamiltonian vector fields.
At the commutative level, this means that we will be able to write actions for
scalars and spinors interacting with gauge fields just (like in (iv))
in terms
of 1) suitable Poisson brackets  2) integration over a suitable
volume form on the spacetime. In this way, we may convert 
such actions
into their noncommutative analogues (like (iv) to (v)).
 These noncommutative actions  will
be the same as their commutative counterparts at large distances, they
will describe  theories possessing the same set of symmetries as in
 the commutative case
 but  containing
only finite number of degrees of freedom! 

 In  chapter 2, we describe
a noncommutative complex of differential forms which underlies the
notion of the gauge field in the noncommutative case.  Then in 
chapter 3 we construct subsequently  scalar and  spinor 
electrodynamics.

\section{Hamiltonian de Rham complex of $S^2$}
\subsection{The commutative case}
Consider an algebra $\A$ of functions in two conjugated bosonic variables
$z,\bz$ and two conjugated fermionic ones $b, \bb$ with the
standard graded commutative multiplication but linearly generated (over $\bc$) 
only by
the functions of the form
\be {\bz^{\bar k} z^k  \bb^{\bar l} b^l\over (1+\bz z +\bb b)^m}, \quad
max(\bar k +\bar l, k+l) \leq m, \quad k,\bar k,l,\bar l,m\geq 0.\ee
 The  algebra is equipped with the {\it graded} involution
\be z^{\*}=\bz, \quad \bz^{\*}=z, \quad b^{\*}=\bb, \quad \bb^{\*}=-b;\ee
\be (AB)^{\*}=(-1)^{AB}B^{\*}A^{\*}, 
\quad (A^{\*})^{\*}=(-1)^{A} A\ee
and it is known as the algebra of functions on the supersphere $CP(1,1)$.
We can define an integral over an element $f$ in the algebra as follows
\be I[f]\equiv
 -{i\over 2\pi}\int {d\bz \w dz \w  d\bb \w db\over 1+\bz z +\bb b} f.\ee
(Note $I[1]=1$.) Now an inner product on $\A$ is defined simply as
\be (f,g)=I(f^{\*}g).\ee
The remaining structure to be defined is the graded symplectic structure,
given by a non-degenerate
 super-Poisson bracket $\{ .,.\}:\A\times\A\to\A$ 
\bea \{f,g\}=(1+\bz z)(1+\bz z + \bb b)(\dz f \dbz g - \dbz f \dz g)\cr
+(-1)^f (1+\bz z)\bb z (\dz f \d_{\bb} g -
\d_{\bb} f \dz g) +(1+\bz z) b\bz (\d_b f \dbz g -
 (-1)^f\dbz f \d_b g)\cr
 +(-1)^{(f+1)}(1+\bz z -\bb b \bz z) (\d_b f \d_{\bb}g + \d_{\bb}f \d_b g).
\eea 
 Now let us introduce four odd vector fields on $\A$:
 \be T_1=\bz \d_{\bb} -b \dz, \quad T_2=\d_{\bb} +zb\dz, \quad \bar T_1
 =\bb \dbz +z\d_b, \quad \bar T_2 =\d_b -\bz \bb \dbz.\ee
 It turns out that $T_i, \bar T_i$ are the Hamiltonian vector fields 
 with respect to the super-Poisson bracket (6). They are  generated by the 
  Hamiltonians $t_i, \bar t_i $:
  \be T_i f= \{t_i, f\}, \quad \bar T_i f = \{\bar t_i, f\};\ee
  \be t_1 \equiv {\bz b\over 1+\bz z +\bb b}, \quad t_2\equiv
   { b\over 1+\bz z +\bb b}, \bar t_1\equiv {\bb z\over 1+\bz z +\bb b},
   \bar t_2\equiv {\bb \over 1+\bz z +\bb b}.\ee
 Note also the 
 properties of $T_i,\bar T_i$  with respect to the graded involution:
 \be (T_i f)^{\*}=-\bar T_i f^{\*}, \quad (\bar T_i f)^{\*} = T_i f^{\*}.\ee
 
 Define a complexified Hamiltonian de Rham complex $\o$  over the 
 standard sphere
 $S^2$ 
 as the graded associative algebra with unit
 \be \o= \o_0 \oplus \o_1 \oplus \o_2,\ee
 where 
 \be \o_0=\o_2=\A_e\ee
 and 
 \be \o_1 =\A_b\oplus \A_b \oplus \A_{\bb}\oplus \A_{\bb}.\ee
 Here $\A_e$ is the even subalgebra of $\A$  linearly generated over $\bc$
 only by
 elements with  $l=\bar l$ (cf. Eq. (1)) and $\A_b$ ($\A_{\bb}$) are (odd)
 bimodules over $\A_e$ linearly generated by the elements of the form (1)
  with $\bar l=0$, $l=1$ ($\bar l=1$, $l=0$). The multiplication in $\o$
  is entailed by one in $\A$, the only non-obvious thing is  to define
  the product of 1-forms. Here it is
  \be (A_1,A_2,\bar A_1,\bar A_2) (B_1,B_2,\bar B_1,\bar B_2)\equiv
  A_1 \bar B_1 +A_2 \bar B_2 +\bar A_1 B_1 +\bar A_2 B_2.\ee
  Of course,  the r.h.s. is viewed as an element
  of $\o_2$.
  The product of a 1-form and a 2-form is set to zero by definition.
  Now the coboundary operator $d$ is given by
  \be df\equiv (T_1 f, T_2 f, \bar T_1 f, \bar T_2 f), f\in \o_0;\ee
  \be d(A_1,A_2,\bar A_1,\bar A_2)\equiv T_1 \bar A_1 +T_2 \bar A_2
  +\bar T_1 A_1 +\bar T_2 A_2, \quad (A_1,A_2,\bar A_1,\bar A_2)\in \o_1;\ee
  \be dh=0, \quad h\in \o_2.\ee
  It maps $\o_i$ to $\o_{i+1}$ and it satisfies
  \be d^2=0, \quad d(AB)= (dA)B +(-1)^A A(dB).\ee
  
  Now we show that the complex $\o$ resembles very much the standard
  (complexified)
  de Rham complex $\o_{dR}$ 
  of the commutative sphere $S^2$ (not of the supersphere!). The latter
  can be defined again as the graded associative algebra with unit given
  by
  \be \o_{dR}=\om_0\oplus\om_1\oplus\om_2,\ee
   where
   \be \om_{0,2}=\B ,\quad \om_1=\B_2 \oplus \B_{\bar 2}.\ee
   Here $\B$ is unital 
   algebra linearly generated (over $\bc$) by elements of the
   form
   \be {\bz^{\bar k} z^k \over (1+\bz z)^m}, \quad
max(\bar k ,k)\leq m, \quad k,\bar k,m \geq 0,\ee
that is, $\B$ is the algebra of complex functions on $S^2$.
 $\B_2$  and $B_{\bar 2}$ are   $\B$-bimodules
  linearly generated by the following elements
of $\B$
  \be {\bz^{\bar k} z^k \over (1+\bz z)^m}, \quad
max(\bar k,k+2) \leq m, \quad k,\bar k, m\geq 0\ee
and 
  \be {\bz^{\bar k} z^k \over (1+\bz z)^m}, \quad
max(\bar k+2,k) \leq m, \quad k,\bar k, m\geq 0,\ee
respectively. 
One often writes the elements $(V,\bar V)$ of $\om_1$
as 
\be   Vdz+\bar V d\bz\ee
and $h$ of $\om_2$ as
\be h {2\over i(\bz z+1)^2} d\bz\w dz, \quad h\in\om_2.\ee
The multiplication in $\o_{dR}$ is now given by the  standard wedge product
in the representation (24,25)
and the de Rham coboundary operator $d_{dR}$ is given in the standard way.

One can easily verify the following
\vskip1pc
\noindent {\bf Fact}: The
complex $\o_{dR}$ can be injected into the complex $\o$. The injection
$\nu$ is a homomorphism which  preserves all involved structures 
(i.e linear, multiplicative and differential ones). It is given explicitly
as follows
\be f(z,\bz)\in\om_0 \to f(z,\bz)\in\o_0;\ee
\be (V(z,\bz),\bar V(z,\bz))\in \om_1\to (-V(z,\bz)b,V(z,\bz)zb,
+\bar V(z,\bz)\bb,-\bar V(z,\bz)\bz\bb)\in\o_1;\ee
\be h(z,\bz) \in \om_2 \to {2i\over (1+\bz z)}h(z,\bz) \bb b \in \o_2.\ee
The injection (26) of $0$-forms requires, perhaps, some explanation.
The elements of $\o_0$ were said to be linearly generated by the
expressions of the form (1) with $l=\bar l$. However, by noting the identity
\be {1\over \bz z+1}\equiv {1\over \bz z +\bb b +1} 
+{\bb b\over (\bz z+\bb b+1)^2}\ee
one can easily see
that any element of $\B\equiv\om_0$ (cf. (21))
can be written as a linear combination of the quantities (1). 

Thus we have injected the standard de Rham complex into the bigger
Hamiltonian de Rham complex
$\o$ which has the virtue that all vector fields (i.e. $T_i,\bar T_i$)
 needed for the definition of the exterior derivative $d$ are now Hamiltonian.
 This means that we have good chance to quantize the structure while
 maintaining all its properties (except  graded commutativity of the 
 multiplication in $\o$). 
 
 There remains to clarify the issues of reality, Hodge star and cohomology.
 As already the name suggests, the complexified Hamiltonian complex
 has a real subcomplex $\o_R$ given by all elements of $\o$ real under
 an involution $\j$ defined as follows
 \be f^{\j}=f^{\*}, ~f\in \o_0, \quad h^{\j}=-h^{\*},~h\in\o_2;\ee
 \be (A_i,\bar A_i)^{\j}=(\bar A_i^{\*},-A_i^{\*}), ~(A_i,\bar A_i)\in \o_1.\ee
 The involution $\j$ ($\j^2=1$) preserves the linear combinations with
 real coefficients and the multiplication,
  and commutes with the coboundary operator
 $d$:
 \be (af+bg)^{\j}=af^{\j} +bg^{\j}, ~a,b\in{\bf R},~f,g\in\o;\ee
 \be (fg)^{\j}=f^{\j}g^{\j},~f,g\in\o;\ee
 \be (df)^{\j}=df^{\j}, f\in\o.\ee
 We see that $\o_R$ is indeed the real subcomplex which may be called
  the Hamiltonian de Rham complex of $S^2$.
 
 We recall that the standard involution $\j$  on $\o_{dR}$ is given by
\be f^{\j}=f^*, ~f\in \om_0, \quad g^{\j}=g^*,~g\in\om_2;\ee
 \be (V,\bar V)^{\j}=(\bar V^*,V^*), ~(V,\bar V)\in \om_1,\ee 
 where $*$ is the standard complex conjugation.
 It has also the property of preserving the real linear combinations and the 
 multiplication in $\o_{dR}$ and it commutes with the standard de Rham
 coboundary operator. Thus the real elements of $\o_{dR}$ form the
 real de Rham complex of $S^2$.
 
 It is now easy thing to check that the homomorphism $\nu: \o_{dR}\to\o$
 preserves the involution, i.e.
 \be (\nu(f))^{\j}=\nu(f^{\j}).\ee
 This is also the reason, why we have chosen the same symbol
  for both involutions.
 
It is instructive to compute the cohomology of the both real complexes:

\noindent 1)The standard de Rham case.

\noindent The only non-trivial class occurs in the second cohomology and
it is given by an element $1$ in $\om_2$

\noindent 2)The Hamiltonian de Rham case: 

\noindent  Using the homomorphism $\nu$, 
the de Rham class above  can be injected into $H^*(\o_R)$. It is
not very difficult to verify, that this is the only non-trivial class there.
\newpage

\noindent {\bf The Hodge star and the inner product}:

  \vskip1pc
  
The Hodge star $*_H$ on the standard de Rham complex $\o_{dR}$ is given
by
\be *_H: ~f(z,\bz)\in\om_0\to f(z,\bz)\in\om_2, \quad h(z\bz)\in\om_2 \to 
h(z,\bz)\in\om_0;\ee
\be *_H: ~(V,\bar V)\in\om_1\to (iV,-i\bar V)\in\om_1.\ee
Note that $*_H$ send  real forms into  real ones. 
The Hodge star $*$ on the Hamiltonian de Rham complex is given  by
\be *: ~f(\bz,z,\bb, b)\in\o_0\to 
 {2if(\bz,z,\bb, b)\bb b\over \bz z+1}\in\o_2;
\ee
\be *: ~(A_1,A_2,\bar A_1,\bar A_2)\in\o_1\to (iA_1,iA_2,-i\bar A_1,-i\bar A_2)
\in\o_1;\ee
\be *: ~h(\bz,z,\bb,b)\in\o_2\to 
{1\over 4i}(\bar T_i T_i -T_i \bar T_i-2)h(\bz,z,\bb,b)\in\o_0.\ee
This Hodge star is also compatible with the involution $\j$ and 
 has the property
\be *\nu(f)=\nu(*_H f).\ee
 This in turn means, that the natural inner product
in $\o_{dR}$
\be (X,Y)_{dR}\equiv {1\over 4\pi}\int (*_HX^{\j}) Y,\quad X,Y\in 
\om_0,\om_1,\om_2\ee
does respect the natural inner product in $\o$
\be (X',Y')\equiv {i\over 2}I[(*X'^{\j}) Y'], \quad X',Y'\in \o_0,\o_1,\o_2.\ee
In other words:
\be (\nu(X),\nu(Y))=(X,Y)_{dR}.\ee
It should be perhaps noted, for clarity, that in (45) and in all the rest
of the paper the integral $I$ is applied {\it always} on element of $\A_e$.
Though in (45) the argument of $I$ is always a $2$-form (to be understood
as an element of $\A_e$) in subsequent applications we shall encounter also
situations in which the argument will be a $0$-form.
 \newpage
 \noindent{\bf The action of $SU(2)$ }
 \vskip1pc 
The standard action of the group $SU(2)$ on $S^2$ induces the action
of the same group on the Hamiltonian de Rham complex.  The latter
respects the grading of the complex; on the $0$-forms (from $\o_0\equiv\A_e$)
and $2$-forms (from $\o_2\equiv\A_e$)
 it is given by  even Hamiltonian vector
field $R_{\pm}, R_3$ obtained by taking suitable anticommutators of the
odd vector fields $T_j,\bar T_j$:
\be R_+=[T_1,\bar T_2]_+=-\partial_z-\bz^2\partial_{\bz}-
\bz \bb \partial_{\bb};\ee
\be R_-=[T_2,\bar T_1]_+=\partial_{\bz}+z^2\partial_z+zb \partial_b;\ee
\be R_3={1\over 2}
([T_1,\bar T_1]_+ -[T_2,\bar T_2]_+)=\bz\partial_{\bz}-z\partial_z+{1\over 2}
\bb\partial_{\bb}-{1\over 2}b\partial b.\ee
The $SU(2)$ Lie algebra commutation relations
\be [R_3,R_{\pm}]=\pm R_{\pm},\quad [R_+,R_-]=2R_3\ee
then directly follows. The Hamiltonians $r_j$ of the vector fields $R_j$ are
obtained by taking the corresponding Poisson brackets (6) of the 
Hamiltonians $t_i,\bar t_i$ given in (9).

The vector fields $R_j$ acting on the algebra $\A$ realize a (highly reducible)
 representation of $SU(2)$. This representation is unitary with respect to the
 inner product (5) and the representation space $\A$ has several invariant
 subspaces which are of interest for us. They are $\A_e$, $\A_b$ 
 and $\A_{\bb}$;
 all of them give rise to smaller unitary representations of $SU(2)$ than $\A$.
 In particular, since both $\o_0$ and $\o_2$ can be identified with $\A_e$,
 we have an action of $SU(2)$ on the $0$-forms and $2$-forms of the
 Hamiltonian de Rham complex $\o$.

Now we realize that the space $\o_1$ of the Hamiltonian 1-forms can be written
as 
\be \o_1=\A_b\otimes \bc^2\oplus \A_{\bb}\otimes\bc^2.\ee
The group $SU(2)$ can be represented on the second copy of 
$\bc^2$ in (51) by the standard spin 1/2
representation generated by the Pauli matrices
\be \si^+=\left(\matrix{0&1\cr 0&0}\right),\quad 
\si^-=\left(\matrix{0&0\cr 1&0}\right),\quad 
\si^3=\left(\matrix{1&0\cr 0&-1}\right)\ee
and on the first copy of $\bc^2$ 
 by its (equivalent)  complex conjugated representation ($\sigma^{\pm}\to 
 -\sigma^{\mp}, ~~\si^3\to -\si^3$).
Now we can define the action of $SU(2)$ on the 1-forms from $\o_1$
again by the formula (51) where $\oplus$ and $\otimes$ are understood
to be the direct product and the direct sum
of  the $SU(2)$ representations.

Now it is easy to check
that 

\noindent 1) the standard de Rham complex $\o_{dR}$  injected in $\o$ 
by the homomorphism $\nu$ is
also an invariant subspace 
of the just defined $SU(2)$ action on $\o$;

\noindent 2) the $SU(2)$ 
action on $\o$
preserves the inner product (45)
and restricted to the image of $\nu$ gives the standard (unitary) action
of $SU(2)$ on the de Rahm complex $\o_{dR}$;

\noindent 3) the coboundary operator $d$ of $\o$ and the Hodge star $*$
are both $SU(2)$ invariant.

\noindent 4) the generators $R_i$ of $SU(2)$ act on $\o$ as derivations
with respect to the product on $\o$, i.e. 
$R_i(\Phi\Psi)=(R_i\Phi)\Psi+\Phi R_i\Psi$.
 
\subsection{The non-commutative case.}
In the previous section, we have described the Hamiltonian de Rham
complex, by using substantially the structure of the algebra of functions
on the supersphere (1). This algebra can be described in an alternative way,
which makes the explicit form of the Poisson structure (6) much less
cumbersome, though it makes more involved the relation between the
standard de Rham complex and its Hamiltonian counterpart. Here are the
details:

Consider the algebra of functions on the
complex $C^{2,1}$ superplane, i.e. algebra generated by  bosonic variables
$\bar\chi^i,\chi^i, i=1,2$ and by fermionic ones $\bar a,a$.
The algebra is equipped with the graded involution
\be (\chi^i)^{\*}=\bar\chi^i,\quad, (\bar\chi^i)^{\*}=\chi^i,\quad, 
a^{\*}=\bar a,
\quad, \bar a^{\*}=-a\ee
and with the super-Poisson bracket 
\be \{f,g\}=\d_{\chi^i}f\d_{\bar\chi^i}g-\d_{\bar\chi^i}f\d_{\chi^i}g
+(-1)^{f+1}[\d_a f\d_{\bar a}g+\d_{\bar a} f\d_a g].\ee
Here and in what follows, the Einstein summation convention applies.
We can now apply the (super)symplectic reduction with respect to a
moment map $\bar\chi^i\chi^i +\bar a a $.  
The result is a smaller algebra
$\A$, that by definition consists of all functions $f$ with the property
\be \{f,\bar\chi^i\chi^i+\bar a a\}=0.\ee
Moreover, two functions obeying (55) are considered to be equivalent
if they differ just by a product
 of $(\bar\chi^i\chi^i+\bar a a-1)$ with some other such function.
The algebra $\A$ is just the same algebra (1) that we have considered in the
previous section. The Poisson bracket (6) becomes the bracket (54) for the
functions in $\A$. The relation between the generators is as follows
\be z={\chi^1\over\chi^2}, \quad \bar z ={\bar\chi^1\over\bar\chi^2},
\quad b={a\over \chi^2},\quad \bar b={\bar a\over \bar \chi^2}.\ee
The integral (4) can be written as
\be I[f]=-{1\over 4\pi^2}\int d\bar\chi^1\w d\chi^1\w d\bar\chi^2\w d\chi^2 
\w d\bar a \w da ~\delta(\bar\chi^i \chi^i +\bar a a-1) f.\ee
The vector fields $T_i,\bar T_i$ turn out to be
\be T_i=\bar\chi^i\d_{\bar a}-a\d_{\chi^i},\quad \bar T_i=\bar a\d_{\bar\chi^i}
+\chi^i\d_a.\ee
Of course, they annihilate the moment map $(\bar\chi^i\chi^i +\bar a a)$,
otherwise they would not be well defined differential operators acting on $\A$.
Their Hamiltonians are
\be t_i =\bar\chi^i a,\quad \bar t_i=\chi^i \bar a.\ee

Now we are ready to quantize the infinitely dimensional
algebra  $\A$ with the goal of obtaining 
its (noncommutative) finite dimensional deformation. 
The quantization was actually performed in \cite{GKP2} using the representation
theory of $osp(2,2)$ superalgebra. Here we adopt a different procedure,
namely the quantum symplectic reduction (or, in other words, quantization
with constraints). This method should be more transparent for anybody who knows
the elements of quantum mechanics.
We start with the well-known quantization of the complex plane $C^{2,1}$.
The generators $\bchi,\chi^i,\ba, a$ become creation and annihilation
operators on the Fock space whose commutation relations are given
by the standard replacement
\be \{.,.\}\to {1\over h}[.,.].\ee
Here $h$ is a real parameter (we have absorbed the imaginary unit into
the definition of the Poisson bracket) referred to as the "Planck constant".
Explicitely
\be [\chi^i,\bar\chi^j]_-=h\delta^{ij}, \quad [a,\bar a]_+=h\ee
and all remaining graded commutators vanish. The Fock space is built up
as usual, applying the creation operators $\bchi,\ba$ on the vacuum $\vert 0
\rangle$, which is in turn annihilated by the annihilation operators $\chi,a$.
The scalar product on the Fock space is fixed by the requirement
that the barred generators are adjoint of the unbarred ones.
We hope that  using the same symbols for the classical and quantum generators
will not confuse the reader; it should be fairly obvious from the context
which usage we have in mind.

Now we perform  the quantum symplectic reduction
 with the self-adjoint moment map
$ (\bchi\chi^i +\ba a)$. First we restrict the Hilbert
space only to the vectors $\psi$ satisfying the constraint
\be (\bchi\chi^i +\ba a -1)\psi=0.\ee
Hence operators $\hat f$ acting on this restricted  space have  to fulfil
\be [\hat f,\bchi\chi^i +\ba a]=0\ee
and they are to form our deformed version\footnote{Note that (63) 
is just a quantum
version of (55) and it says that elements of the deformed algebra 
have to  commute with the particle number operator.} of $\A$.

The spectrum of the operator $(\bchi\chi^i+\ba a -1)$ in the
Fock space is given by a sequence $mh-1$, where $m$'s are integers.  In order
to fulfil (62) for a non-vanishing $\psi$, we observe that the 
inverse Planck constant
$1/h$ must be an integer $N$. The constraint (62) then selects only
$\psi$'s living in the eigenspace $H_N$ 
of the operator $(\bchi\chi^i +\ba a -1)$
with the eigenvalue $0$. This subspace of the   Fock space
has the dimension $2N+1$ and the
algebra $\A_N$ of operators $\hat f$ acting on it is $(2N+1)^2$-dimensional.

 When
$N\to\infty$ (the dimension $(2N+1)^2$ then also diverges) we have 
the Planck constant approaching $0$ and, hence, the algebras $\A_N$
tend to the classical limit $\A$.
 The fact that the resulting finite-dimensional noncommutative
algebras $\A_N$ are  deformations of $\A$ is thus  clear since
the latter is just the classical limit of the former. The interested
reader may find a rigorous proof of this fact in \cite{GKP2}.

The Hilbert space $H_N$ is naturally graded. The even subspace $H_{eN}$
is  created from the Fock vacuum by applying only the bosonic creation
operators:
\be (\bar\chi^1)^{n_1}(\bar\chi^2)^{n_2}\vert 0\rangle,\quad n_1+n_2=N,\ee
while the odd one $H_{oN}$ by applying both bosonic and fermionic creation
operators:
\be (\bar\chi^1)^{n_1}(\bar\chi^2)^{n_2} \ba\vert 0\rangle, \quad n_1+n_2 
 =N-1.\ee
Correspondingly, the algebra of operators $\A_N$ on $H_N$ consists of
 an even part $\A_{eN}$
(operators respecting the grading) and an odd part (operators reversing
the grading). The odd part can be itself written as a direct sum
$\A_{aN}\oplus\A_{\bar a N}$. The two components in the sum are distinguished
by their images: $\A_{aN}H_N=H_{eN}$ while $\A_{\bar aN}H_N=H_{oN}$.
The algebra $\A_{aN}$ is spaned by operators
\be (\bar\chi^1)^{n_1}(\bar\chi^2)^{n_2}(\chi^1)^{m_1}(\chi^2)^{m_2}a,\quad 
n_1+n_2=m_1+m_2 +1=N,\ee
 $\A_{\bar aN}$ by
\be (\bar\chi^1)^{n_1}(\bar\chi^2)^{n_2}\ba (\chi^1)^{m_1}(\chi^2)^{m_2},\quad 
n_1+n_2+1=m_1+m_2=N\ee
and $\A_{eN}$ by
\be (\bar\chi^1)^{n_1}(\bar\chi^2)^{n_2}(\chi^1)^{m_1}(\chi^2)^{m_2}
(\ba a)^k,\quad 
n_1+n_2=m_1+m_2 =N-k.\ee
From this and (56),
 it is obvious that $\A_e$ from the previous section got deformed
to $\A_{eN}$, while $\A_b$ and $\A_{\bb}$ to $\A_{aN}$ and $\A_{\ba N}$,
respectively. 

The inner product (5) on $\A$ is given by the integral $I$ (4) or (57).
Its representation (57) is more convenient for finding its noncommutative
deformation. At the level of supercomplex plane $C^{2,1}$ 
it is the textbook fact
from quantum mechanics that the integral  $\int d\bchi d\chi^i d\ba da$
(this is the Liouville integral over the superphase space)
is replaced under the quantization procedure 
by the supertrace in the Fock space. (The 
supertrace is the trace over 
the indices of the zero-fermion states minus the trace over the one-fermion
states). The $\delta$ function of the
operator $(\bchi\chi^i +\ba a-1)$ just restrict the supertrace to the
trace over the indices of
$H_{eN}$ minus the trace over the indices of $H_{oN}$. Hence
\be (\hat f, \hat g)_N
\equiv \ST [\hat f^{\*} \hat g],\quad \hat f,\hat g\in\A_N.\ee
  Here the graded involution $\*$  in the noncommutative algebra $\A_N$
 is defined exactly as in (53).
It is now obvious that this
 inner product approaches in the limit $N\to\infty$ the commutative one.
This detailed proof of this fact was furnished in \cite{GKP2}.

Define a non-commutative Hamiltonian de Rham complex $\o_N$ of the 
fuzzy sphere $S^2$ as the graded associative algebra with unit
\be \o_N=\o_{0N} \oplus \o_{1N}\oplus \o_{2N},\ee
where 
\be \o_{0N}=\o_{2N}=\A_{eN}\ee
and 
\be \o_{1N}=\A_{aN}\oplus\A_{aN}\oplus\A_{\ba N}\oplus\A_{\ba N}.\ee
The multiplication in $\o_N$
  with the standard properties with respect to the grading
  is entailed by one in $\A_N$. The product of 1-forms
   is given by the same formula as in the
  graded commutative case  (14)
  \be (A_1,A_2,\bar A_1,\bar A_2) (B_1,B_2,\bar B_1,\bar B_2)\equiv
  A_1 \bar B_1 +A_2 \bar B_2 +\bar A_1 B_1 +\bar A_2 B_2.\ee
  Of course, by definition, the r.h.s. is viewed as an element
  of $\o_{2N}$. Here we note an  important difference with the graded
  commutative case: the product $AA$ of a 1-form $A$ with itself automatically
  vanishes in the commutative case but may be a non-vanishing element of 
  $\o_{2N}$ in the deformed picture.
  The product of a 1-form and a 2-form is again set to zero by definition.
  Now the coboundary operator $d$ is given by
  \be df\equiv (T_1 f, T_2 f, \bar T_1 f, \bar T_2 f),\quad f\in \o_{0N};\ee
  \be d(A_1,A_2,\bar A_1,\bar A_2)\equiv T_1 \bar A_1 +T_2 \bar A_2
  +\bar T_1 A_1 +\bar T_2 A_2, \quad (A_1,A_2,\bar A_1,\bar A_2)\in \o_{1N};\ee
  \be dh=0, \quad h\in \o_{2N},\ee
  where the action of $T_i,\bar T_i$ is given by the noncommutative version
  of (8):
  \be T_i X\equiv N(t_i X -(-1)^X X t_i), \quad \bar T_i X\equiv
   N(\bar t_i X -(-1)^X X \bar t_i), \quad X\in \A_N,\ee
   where 
   \be \quad
    t_i =\bar\chi^i a,\quad \bar t_i=\chi^i \bar a.\ee
$d$ maps $\o_{iN}$ to $\o_{i+1,N}$ and it satisfies
  \be d^2=0, \quad d(AB)= (dA)B +(-1)^A A(dB).\ee
 Using the graded involution $\*$, we define the standard
involution $\j( \j^2=1)$  on the noncommutative complex $\o_N$:
 \be f^{\j}=f^{\*}, ~f\in \o_{0N}, \quad g^{\j}=-g^{\*},~g\in\o_{2N};\ee
 \be (A_i,\bar A_i)^{\j}=(\bar A_i^{\*},-A_i^{\*}), ~(A_i,\bar A_i)\in \o_{1N}.
 \ee
 The coboundary map $d$ is compatible with the involution, however,
  due to noncommutativity,
  it is no longer true that the product of two real elements of
 $\o_N$ gives a real element. Thus we cannot define the real noncommutative
 Hamiltonian de Rham complex. For field theoretical applications
 this is not a drawback, nevertheless, because for the formulation
 of the field theories we shall not need the structure of the real subcomplex,
 but only the involution on the complex Hamiltonian de Rham complex.

 The Hodge star $*$ in $\o_N$ is defined precisely
 as in the commutative case:
 
\be *: ~f\in\o_{0N}\to  2i\bar a af\in\o_{2N};\ee
\be *: ~(A_1,A_2,\bar A_1,\bar A_2)\in\o_{1N}
\to (iA_1,iA_2,-i\bar A_1,-i\bar A_2)
\in\o_{1N};\ee
\be *: ~ h\in \o_{2N}\to 
{1\over 4i}(\bar T_i T_i -T_i\bar T_i-2)h\in\o_{0N}.\ee
This Hodge star is  compatible with the involution $\j$. Note that the
definition of the Hodge star $*$ on the $0$-forms does not involve any
ordering problem, since the operator $\bar aa$ commutes with the elements
of $\o_{0N}$.

The natural inner product on $\o_N$, whose commutative limit is (45),
is 
\be (X,Y)_N\equiv {i\over 2}\ST[(*X^{\j}) Y],\quad X,Y\in 
\o_{N0},\o_{N1},\o_{N2},\ee
where $\ST$ is the standard supertrace (cf. (69)).

\vskip1pc
\noindent{\bf The action of $SU(2)$}
\vskip1pc

The study of the $SU(2)$ action on the deformed
Hamiltonian de Rham complex $\o_N$  is important
in view of our field theoretical applications. We require that such a 
$SU(2)$ action gives in the commutative limit the $SU(2)$ action
on the undeformed de Rham complex, described in the previous section.
This can be easily arranged, however, because the action on the undeformed 
commutative complex is generated
entirely in terms of the Hamiltonian vector fields  $R_j$ whose
Hamiltonians are $r_j$:
\be r_+=\chi_1^{\j}\chi_2, \quad r_-=\chi_2^{\j}\chi_1, \quad r_3=
{1\over 2}(\chi_1^{\j}\chi_1 -\chi_2^{\j}\chi_2).\ee
Hence the deformed action on $\A_{N}$ will be generated by the same
Hamiltonians (86) (now understood as  operators on the Fock space)
but the Poisson bracket will be replaced by the commutator:
\be R_j X =N[r_j,X],\quad X\in\A_N.\ee
It is trivial to check that the commutation relations (50) are fulfilled
for this definition
and that the $SU(2)$ representation so generated is unitary 
with respect to the deformed inner product (69) on $\A_N$.
Since both $\o_{0N}$ and $\o_{2N}$ can be identified with $\A_{eN}$, 
which is (as in the nondeformed case) an invariant subspace
of $\A_N$,  we have
just obtained the $SU(2)$ action on the even forms of the deformed
Hamiltonian complex $\o_N$. Recall that the space $\o_{1N}$ of the 1-forms
can be written as \be
\o_1=\A_{a N}\otimes \bc^2\oplus \A_{\bar a N}\otimes\bc^2.\ee
As in the nondeformed case, the spaces $\A_{aN}$ and $\A_{\bar aN}$
are invariant subspaces of the $SU(2)$ action on $\A_N$, given by (87), 
thus the
formula (88) makes sense at the level of representations of $SU(2)$.
In other words, $\oplus$ and $\otimes$ are operations on the $SU(2)$
representations thus defining $\o_{1N}$ as a  $SU(2)$ representations
which we look for. (As in the nondeformed case (51),
 the $SU(2)$ acts by  (52) on the second copy of $\bc^2$ in (88)
 and on the first copy it acts in the
 complex conjugated way.)

The final three facts, we shall need, read:  1) the inner product (85)
 on $\o_N$
 is invariant with respect to the just defined $SU(2)$ action; 
 2) the coboundary operator $d$ on $\o_N$ and the Hodge star $*$
 are both $SU(2)$ invariant; 3) the generators $R_i$ of $SU(2)$ act on $\o_N$
 as derivations with respect to the deformed product on $\o_N$.
 
\vskip1pc

We have constructed the non-commutative deformation of the 
complexified commutative Hamiltonian de Rham complex. We have identified
the noncommutative counterparts of all structures of the latter and shown
that they have the correct commutative limit.
In particular, we observed that the multiplication in $\o_N$ approaches 
for $N\to\infty$
the standard commutative product in $\o$.
 We have described the involution, the inner product,
  the Hodge star and the $SU(2)$ action  on $\o_N$ which have also
the correct commutative limits and we have established that the "$d$" in the
noncommutative context has all the basic properties (79) in order to deserve
to be called the coboundary operator. Thus, whatever 
commutative construction which
we perform by using these structures can be rewritten in the deformed
 finite-dimensional case. In particular, we shall write the field
 theoretical actions in this way. 
 
 Note that we did not inject any smaller deformed de Rham complex
 into the deformed Hamiltonian complex .
If we could do this we would not have had to bother ourselves
with the Hamiltonian
case! The point of our construction is that we will be able to
formulate the dynamics
of the standard gauge theories using the commutative
{\it Hamiltonian} complex; this means that
we also can deform those theories directly at this level.
The price to pay is relatively
low: few auxiliary fields will appear on the top of the  fields
present in the more standard constructions.
 We are here in a similar position as people who introduce
auxiliary fields in  trying to achieve a closure of algebras
of supersymmetry without imposing the equations of motion.
It turns out that our auxiliary
fields are just the same as those used for the closure of the algebra of
 supersymmetry! Thus the same auxiliary fields do the double job. We believe
that this is not just a coincidence.

\vskip1pc

\noindent {\bf Noncommutative Poincar\'e lemma and cohomology}

\vskip1pc
Although the following few lines will be of no immediate use for our
discussion of the field theoretical actions in this article, it is
interesting to remark, that the
finite deformed complex $\o_N$ preserves faithfully the cohomological
content of $\o$ as the following noncommutative generalization of the
 Poincar\'e
lemma says:
\vskip1pc
\noindent {\bf Theorem}:
\vskip1pc
\noindent i) Let $f\in\o_{0N}, \quad df=0$. Then $f$ is the 
unit element of $\o_{eN}$ (unit matrix acting  on $H_N$)
multiplied by some complex number.

\noindent ii) Let $A\equiv(A_1,A_2,\bar A_1,\bar A_2)\in\o_{1N}, \quad dA=0,
\quad A=A^{\j}$.
Then $A=dg$ for some $g\in\o_{0N},~g=g^{\j}$.

\noindent iii) Let $F\in\o_{2N}$ (i.e. $dF$  automatically vanishes),
$F=F^{\j}$. Then
$F$ can be written as $F=pId+dB$, where $B\in\o_{1N}, ~B=B^{\j}$ 
is some 1-form,
 $Id$
is the unit element in $\A_{eN}$ and $p$ is an imaginary number.
$Id\in\o_{2N}$ itself cannot be written as a coboundary of some 1-form.

\vskip1pc  
Thus the theorem implies that 
$Id$ is the only nontrivial cohomology class in $H^2(\o_N)$ and $H^0(\o_N)$,
 and $H^1(\o_N)$
vanishes.
\vskip1pc
\noindent {\bf Proof}:
\vskip1pc
\noindent i) One notices that the Hamiltonians $t_i,\bar t_i$
of the vector fields $T_i,\bar T_i$ generates the whole algebra $\A_N$ and
therefore also its subalgebra $\A_{eN}=\o_{0N}$. According to (77), the
$T_i,\bar T_i$ act on an element $f\in\o_{0N}$ as commutators $N[t_i,f],
N[\bar t_i,f]$, respectively.
 Thus vanishing of the commutators means that $f$ commutes with
all matrices in $\A_N$. Hence $f$ is a multiple of the unit matrix $Id$.

\noindent ii) It is easy to see that an  arbitrary real 1-form $A\in\o_{1N}$ 
has  $2N(N+1)$ complex components. The condition $dA=0$ sets
on them $(N+1)^2+N^2-1$ real independent constraints (the term $-1$ comes
 from the fact that $\ST (dA)\equiv 0$
for every $A$).  Thus a solution $A=A^{\j}$
 of this condition must depend on 
 \be 2N(N+1)=(N+1)^2 +N^2 -1\ee
  real parametres.
 The explicit form of the equation $dA=0$ can be written by using the
 matrices $t_i,\bar t_i$:
 \be (t_1)_{jk}=\sqrt{N-j+1}\delta_{jk},
 \quad j=1,\dots,N+1; ~k=1,\dots, N;\ee
 \be (t_2)_{jk}=\sqrt{j-1}\delta_{j-1,k},\quad j=1,\dots,N+1; ~k=1,\dots, N;\ee
 \be (\bar t_1)_{jk}=\sqrt{N-j+1}\delta_{jk},
 \quad j=1,\dots,N; ~k=1,\dots, N+1;\ee
 \be (\bar t_2)_{jk}=\sqrt{j}\delta_{j+1,k},
 \quad j=1,\dots,N; ~k=1,\dots, N+1.\ee
 Now the direct inspection reveals that indeed the remaining $(N+1)^2+N^2-1$
 real parameters organizes into one real element $g=g^{\j}$
 of $\A_{0N}=\o_{0N}$ ($A=dg$). Note a small redundancy, however: 
  If some $g'$ differs from $g$ by a real multiple of $Id$,
 it gives the same $A$; the term $(-1)$ in (89) precisely corresponds to this
 (cohomological) ambiguity. This means, for example, that the solution $g$
 ($dg=A$) can be chosen to have a vanishing supertrace, i.e. it really
 has precisely $(N+1)^2 +N^2-1$ free real parametres.
 
 \noindent  iii) We can do a similar counting as in ii). Again, an
 arbitrary real 1-form $A\in\o_{1N}$ 
has in total $2N(N+1)$ complex components. If we consider  varying $A$,
then  $dA$ sweeps a submanifold $S$ in the manifold of 
real 2-forms. The real dimension of this 
submanifold $S$ cannot exceed $(N+1)^2+N^2-1$.
 The counting is easy: since $A=df$ is in the
kernel of $dA$, varying of $(N+1)^2+N^2-1=2N(N+1)$  real parameters 
of $f$ does not show up in $dA$. There remains
 $(N+1)^2 +N^2-1$ real parameters to vary in $S$.
 
  Now
 a real dimension of the whole manifold of the real two forms is $(N+1)^2+N^2$,
 i.e. bigger by a factor $1$. Using the explicit form (90-93) of $t_i,\bar t_i$,
 it is straightforward to check that $S$ consists of all
  real 2-forms in $\o_{2N}$ with  a vanishing supertrace. Thus the real
   dimension of $S$ is, in fact, precisely $(N+1)^2 +N^2 -1$.  The only
    nontrivial
   cohomology class in $H^2(\o_N)$ is therefore supertraceful and can
   be chosen to be an imaginary multiple of $Id$. The theorem is proved.
   
\section{Gauge theories on noncommutative $S^2$}
\subsection{Theories with a scalar matter.}
Consider a complex scalar field $\phi$ on $S^2$ (i.e. $\phi\in\B$; cf. (21))
and an $U(1)$ gauge connection
 described on the complement of the north (south) pole by a  
real 1-form field 
$v^N=V^N(\bz ,z)dz+ V^{*N}(\bz,z) d\bz$ ($v^S=V^S(\bar w,w)dw +V^{*S}(\bar w,w)
 d\bar w$)
where $z,w$ ($z=1/w$) are the complex
 coordinates on  the corresponding patches. In what follows, we understand
 $\phi$ to be always a section of  the trivial line bundle on $S^2$, i.e.
 the standard complex function. Thus we can describe the connection
 by one globally defined $1$-form on $S^2$.
  Here we shall  work only with one patch -the complement
 of the north pole.
 We encode  the global character of the 1-form $v$
 \be v=Vdz +  V^* d\bz\ee on the
 patch by demanding that $V(V^*)$ is an element of a $\B$-bimodule 
 $\B_2(\B_{\bar 2})$ (cf. (22,23)).
This ensures that $z^2V$ does not diverge for $z\to\infty$ and, thus, the
form $v$ is well defined globaly over the whole sphere $S^2$.

The scalar electrodynamics is defined by an action
\bea S={1\over 4\pi i}\int d\bz \w dz
 \{(\partial_{\bz}+i V^*)\phi^*(\d_z -iV)\phi+ 
 (\partial_{\bz}-i V^*)\phi(\d_z +iV)\phi^*\cr
-{1\over 2g^2} (1+\bz z)^2 (\d_{\bz}V-\d_z V^*)^2\}.\eea
Here $g$ is a real coupling constant of the theory. As usual, the kinetic
term of the scalar field in two dimension does not "remember" the conformal
factor of the round metric $ds^2$ on the sphere 
(i.e. $ds^2=4d\bz dz (1+\bz z)^{-2}$) 
but the kinetic term of the gauge fields must be multiplied by the
inverse power of the conformal factor. 

The action (95) can be written in the language of  forms (elements
of the standard de Rham complex) as follows
\be S=(d\phi -iv\phi,d\phi-iv\phi)_{dR}+{1\over g^2}(dv,dv)_{dR},\ee
where the inner product $(.,.)_{dR}$ 
was defined in (44).
 
Consider now a theory whose multiplet of fields is given by 
a complex Hamiltonian $0$-form 
 $\Phi\in \A_e$ (cf. (12))  and by a real Hamiltonian 1-form $A\in\o_1$
 ($A=(A_1,A_2,\bar A_1,\bar A_2)$, $A^{\j}=A$) and whose action is given
 by
 \be S_{\infty}=(d\Phi -iA\Phi,d\Phi-iA\Phi) +{1\over g^2}(dA,dA).\ee
 Here the inner product $(.,.)$ on $\o$ was defined in (45).
 (The index $\infty$ refers to the fact that this action  will be soon 
  recovered
 as an $N\to\infty$ limit of an action $S_N$, defined on the deformed
 Hamiltonian complex.)
 In order to understand the content of this theory, let us 
 parametrize the Hamiltonian forms $\Phi$ and $A$ as
 \be \Phi=\phi + {F\bb b\over \bz z+1};\ee
 \bea A_1=-(V+{P\bz\over \bz z+1})b, \quad A_2=(Vz-{P\over \bz z+1})b, \cr
 \bar A_1=(V^*+{P^*z\over \bz z+1})\bb, 
 \quad \bar A_2=-(V^*\bz -{P^*\over \bz z+1})\bb,\eea
where all the fields $\phi, F, V$ and $P$ depend only on the variables
$\bz,z$. It is easy to verify that if $\phi, F$ and $P$ belong to the
algebra $\B$ (cf. (21)) and $V$ and $V^*$ to the bimodules $\B_2$ and 
$\B_{\bar 2}$ respectively, then $A$ given by (99) sweeps the space $\o_1$.
In other words, the multiplet consisting of the 
 complex Hamiltonian $0$-form $\Phi$ and the 
real Hamiltonian $1$-form $A$ contains three standard complex de Rham $0$-forms
(the scalar fields $\phi,P,F$) and one real de Rham $1$-form $v=Vdz+V^*d\bz$.
Moreover note, that the $1$-form $v$ is injected in $A$ by using the 
homomorphism $\nu$ (cf.(27)).

 By using the ansatz (98) and (99), we first compute
 \be dA=(P^*-P)+{\bb b\over 1+\bz z}[(\d_zV^*-\d_{\bz}V)(1+\bz z)^2
 +(P^*-P)]\ee
 and
 \be *dA= {i\over 2}(1+\bz z)^2[(\d_{\bz}V-\d_z V^*)+
 {\bb b\over \bz z+1}~ \d_{\bz}\d_ z
 (P^*-P)]
 \ee
 and  then for the action (97) we have 
 \bea S_{\infty}=(d\phi-iv\phi,d\phi-iv\phi)_{dR}+{1\over g^2}(dv,dv)_{dR}\cr
 {1\over 4g^2}(d(P^*-P),d(P^*-P))_{dR} +{1\over 4}
 ((P^*-P)\phi,(P^*-P)\phi)_{dR}\cr+
 {1\over 4}(2F+i(P^*+P)\phi,2F+i(P^*+P)\phi)_{dR}.\eea
 Here the 
  the combination $2F+i(P^*+P)\phi$   plays the role
 of an auxiliary filed and therefore the last line in (102)
 can be obviously dropped out.

We observe that the 
 result (102) is not quite the  action (96) of the scalar 
electrodynamics but, apart from $\phi$ and $v$,
there is one more propagating interacting
field present, namely the imaginary part of $P$. This new field is neutral
and it couples to the field $\phi$ only. It is not difficult to recognize,
that the action (102) describes nothing but the bosonic sector of the
supersymmetric extension of the Schwinger model \cite{Fer}.
Thus, we have a good chance that upon adding fermions to our framework
 we shall recover
the whole supersymmetric theory! 

It is not difficult to construct also the "pure" scalar electrodynamics (96)
in our approach. For doing this, we have to note that, 
 except of applying the Hodge
star *, there is another way how to convert the $2$-form $dA\in\o_2$ into a 
$0$-form from $\o_0$.
Indeed, since both $\o_0$ and $\o_2$ can be identified with $\A_e$, the
identity map does the job. In what follows, we shall understand by the
symbol $(dA)_0$ the corresponding $0$-form. The standard scalar electrodynamics
is then described by the action
\be S_{\infty}=(d\Phi-iA\Phi,d\Phi-iA\Phi)
-{1\over 4g^2}I[(dA)_0^2]+{i\over 8}I[*(\Phi^{\j}(dA)_0^2\Phi)].\ee
Indeed, apart from the combination $2F+i(P^*+P)\phi$, also the field
$(P^*-P)$ becomes auxiliary and it ought to be integrated away hence giving
the action (96).

It is important to note that the actions (97) and (103) have
 a bigger gauge symmetry
than the standard actions (102) and (96). Indeed, let $G_e$ be a group of 
unitary elements of $\A_e$, i.e.
\be G_e=\{U\in\A_e;~U^{\j}U=UU^{\j}=1\}.\ee
An element $U$ of $G_e$ acts on $(\Phi,A)$ as follows
\be \Phi\to U\Phi, \quad A\to A- i dU~U^{-1}.\ee
With the transformation law (105), 
the actions (97) and (103) are
 gauge invariant with respect to $G_e$. By inspecting
(104), it is not difficult to find that the group $G_e$ decomposes in the
direct product of the standard $U(1)$ gauge  group 
consisting of all elements of the form $e^{i\lambda},\lambda\in\B$
and a $\br$ gauge group (gauged real line)
 which acts only on the
auxiliary fields. This $\br$-subgroup drops out from the formulation involving
only the dynamical fields and we are left with the standard $U(1)$ 
gauge transformations 
\be \phi\to e^{i\lambda}\phi, \quad V\to V+\d_z\lambda, \quad V^*\to V^*
+\d_{\bz}\lambda.\ee

We have rewritten the standard action of the scalar electrodynamics
in terms of the structures of the Hamiltonian de Rham complex. This in turn
means that we can directly write down  a finite dimensional deformation
of the field theoretical model103) by replacing all structures occuring
in (103) by their noncommutative counterparts:
\be S_N=(d\Phi -iA\Phi,d\Phi-iA\Phi)_N -{1\over 4g^2}\ST[(dA-iA^2)_0^2]
+{i\over 8}
\ST[*(\Phi^{\j}(dA-iA^2)^2_0\Phi)].\ee
Here $\Phi\in\o_{0N}$
 is a noncommutative Hamiltonian 0-form, $A\in\o_{1N}$ a real noncommutative
Hamiltonian 1-form and the inner product $(.,.)_N$ is given by (85).

Now we should examine the gauge invariance of this action. Consider
a group $G_{eN}$ consisting of the unitary elements of $\A_{eN}$:
\be G_{eN}=\{U\in\A_{eN};~U^{\j}U=UU^{\j}=1\}.\ee
An element $U$ of $G_{eN}$ acts on $(\Phi,A)\in\o_N$ as follows
\be \Phi\to U\Phi, \quad A\to UAU^{-1}- i dU ~U^{-1}.\ee
Note that due to noncommutativity the  gauge transformation looks like
a non-Abelian one (only in the limit $N\to\infty$ it reduces to the
standard Abelian one). In the action (107), the term $A^2$ 
(which identically vanishes in the
commutative limit)  is crucial for the
gauge invariance for only with it the field strength $dA-iA^2$ transforms
homogenously:
\be dA-iA^2\to U(dA-iA^2)U^{-1}.\ee
It is important to keep in mind that the gauge group $G_{eN}$ is
a deformation of the commutative gauge group $G_e$. The latter 
has its local $U(1)$ subgroup acting as in (106). In the noncommutative
case, we cannot say which part of the Hamiltonian connection $A$ is auxiliary
(i.e. $P$-part)
and which dynamical ($v$-part). Thus neither we can identify a noncommutative 
"local"
$U(1)$ subgroup of $G_{eN}$.  This means that  the full $G_{eN}$ group
plays a  role in the deformed theory. I believe that this fact will be crucial
in getting a nonperturbative insight on the problem of chiral anomaly.

The noncommutative deformation of the model (97) is slightly more
involved than the one of the standard scalar electrodynamics (103) , for
we have to add more terms like $A^2$ which in the commutative limit trivially
vanish but they are required for the gauge invariance of the deformed
model. The easiest way to proceed consists first in rewriting the undeformed
action in the form
\be S_{\infty}=(d\Phi-iA\Phi,d\Phi-iA\Phi)+{1\over 4g^2}(d(dA)_0,d(dA)_0)
-{1\over 4g^2}I[(dA)_0^2].\ee
In order to derive (111) from (97) we have used the explicit form (41,42)
 of the
Hodge star * and the integration "per partes". Now it is easy
to write a deformation of the model (111), which is gauge invariant with
respect to the  transformation laws (109):
\bea S_N=(d\Phi-iA\Phi,d\Phi-iA\Phi)_N-{1\over 4g^2}
\ST[(dA-iA^2)_0^2]\cr+{1\over 4g^2}(d(dA-iA^2)_0
-i[A,(dA-iA^2)_0],d(dA-iA^2)_0-i[A,(dA-iA^2)_0])_N
.\eea
As the term $A^2$, also $i[A,(dA-iA^2)_0]$ vanishes
identically \footnote{It may seem that the commutator gives
a Poisson bracket in the commutative limit but,in fact, 
this is false. The truth is that 
only the commutator multiplied by $N$ gives for $N\to\infty$ the Poisson
bracket.} in the commutative limit, 
thus converting (112) into (111)
for $N\to\infty$.

So far we did not mention a very important property of the
classical scalar electrodynamics on the sphere. Namely, the non-deformed
actions (97) and (103) are invariant with respect to the $SU(2)$ group
 which rotates
the sphere (this $SU(2)$ symmetry is a compact euclidean version
of the standard Poincar\'e symmetry of field theories).  
This statement follows from the invariance of the inner
product (45) on $\o$ (recall that the action of the
group $SU(2)$ on $\A_e$ and on $\o$ was defined in section 2.1),
 the $SU(2)$ invariance of the coboundary operator $d$ and of the Hodge star *
and the fact that the $SU(2)$ generators act as derivations
with respect to the product on $\o$ (i.e. $R_j(\Phi\Psi)=
(R_j\Phi)\Psi +\Phi(R_j\Psi)$).  But all these properties hold also
in the deformed case, thus we conclude that also our deformed actions (107)
and (112)  are
$SU(2)$ invariant.

We end up this section by noting that one can easily formulate theories
with a nontrivial potential energy of the scalar field (like an Abelian
Higgs model) by adding to the  action (103)  
a term of the  form
$(*W(\Phi^{\j}\Phi),1)\equiv {i\over 2}I[*W(\Phi^{\j}\Phi)]$. 
In the deformed case
we have to add to (107) ${i\over 2}\ST[*W(\Phi^{\j}\Phi)]$.
 Here the potential $W$ is some real entire function. After eliminating the
 auxiliary fields in the commutative case, a standard potential 
 term $(*W(\phi^*\phi),1)_{dR}$ turns out  to be added to (96).
In particular, the linear function $W$ corresponds to assigning a mass
to the charged field. 

\subsection{Spinor electrodynamics}

 The standard (chiral or Weyl) spinor bundle on $S^2$ 
 can be identified with the complex line bundle with the winding
number $\pm 1$ (this is to say that the transition function on the overlap
$N\cap S$ of the patches is $({z\over\bz})^{\pm {1\over 2}}=
e^{\pm i\varphi}$ where $\varphi$
is the asimutal angle on the sphere). The plus (minus) sign corresponds to the 
right (left) chirality of the spinor. 
We shall work only with the patch $N$ (the complement
of the north pole) parametrized by the complex
coordinate $z$. A Dirac spinor is then a sum of right and left handed Weyl
spinors:
\be \psi_D=\psi_R +\psi_L.\ee
 Here $\psi_R$ and $\psi_L$ have   each one complex Grassmann valued component
  and 
  they represent  globally well defined sections 
  of the corresponding chiral spinor bundles iff they are 
 elements of a $\B$-bimodules $\ti \B_1$ and $\ti\B_{\bar 1}$ linearly
generated by the elements of the form
\be {\bz^{\bk}z^k\over (1+\bz z)^{m+{1\over 2}}},\quad max(\bar k-1, k)
\leq m,\quad  k,\bar k,m\geq 0\ee 
and
\be {\bz^{\bk}z^k\over (1+\bz z)^{m+{1\over 2}}},\quad max(\bar k, k-1)
\leq m,\quad  k,\bar k,m\geq 0,\ee 
respectively. 
The fact that, say, a right handed spinor $\psi_R$
should be an element of the bimodule (114) follows from the fact
that a spinor bilinear composed of spinors of the same chirality
must be in a line bundle with winding number
$2$. The elements of the latter, multiplied by the zweibein component
\be e_z^u={2\over 1+\bz z}\ee
are therefore components of holomorphic $1$-forms $Vdz$ and we already know
that $V$ must belong to $\B_{2}$ (cf. (22)).
The action of the free euclidean massless Dirac field on $S^2$ is
given by
\be S={i\over 8\pi }\int d\bz \w dz ~e 
~\bar\psi_D \gamma^c e_c^{\mu} (\d_{\mu}
+{1\over 8}\omega_{\mu, ab}[\gamma^a,\gamma^b] )\psi_D,\ee
where $e$ is the determinant of the zweibein $e_{\mu}^a$ 
(or square root of the
determinant of the metric), $\gamma^c$ are flat euclidean 
Hermitian $\gamma$ matrices
satisfying 
\be \{\gamma^a,\gamma^b\}=2\delta^{ab}\ee
and \be 
\bar\psi_D= \bar \psi_R +\bar \psi_L\ee
is the conjugated Dirac spinor ($\bar\psi_{R(L)}$ is now left(right)
handed).

 It is convenient to introduce the flat holomorphic  index $u$
by
\be \gamma^u\equiv\gamma^0 +i\gamma^1=\left(\matrix{0&2\cr 0&0}\right).\ee
Then $\gamma^{\bar u}$ is defined as the Hermitian conjugate of $\gamma^u$.
 The elements of the zweibeins are
 \be e_z^u={2\over 1+\bz z}= e_{\bz}^{\bar u}.\ee
 The last thing to be explained in (117) is the notion of the spin connection
 $\omega_{\mu,ab}$. It is defined by the requirement
 \be \d_{\mu}e_{\nu}^a-\Gamma_{\mu\nu}^{\lambda} e_{\lambda}^a 
 +\omega_{\mu~b}^ae^b_{\nu}=0,\ee
 where $\Gamma_{\mu\nu}^{\lambda}$ are the standard Christoffel symbols.
 For the round metric on the sphere one computes
 \be (\omega_z)_{u\bar u}={1\over 2}{\bz \over 1+\bz z},
 \quad  (\omega_{\bz})_{\bar u u}={1\over 2}{z \over 1+\bz z}.\ee
 It is important to note that the $1$-form $\omega_z dz+\omega _{\bz}d\bz$
  (with values in  the Lie algebra $so(2)\equiv u(1)$) given by (123)
 is not globally defined in the sense of (22,23) since it is singular
 at the north pole $N$ (it is easy to see the singularity by looking
 at $\omega$ in the coordinate patch $w$). 
 This should be the case, however, since $\omega$
 is the connection on the nontrivial spinor bundle. An arbitraty $U(1)$
 connection on the spinor bundle can be achieved by adding a globally
 well-defined form $v=Vdz+V^*d\bz$ (cf. (22,23)) to the spin connection 
 $\omega$.
 Thus,
 the interaction of the Dirac field on $S^2$ with an external $U(1)$
 field $v$ is described by the action
 \be S_v={i\over 2\pi}
 \int {d\bz \w dz \over \bz z+1}\{\bar\psi_R(\d_z-{1\over 2}
 {\bz \over 1+\bz z}-iV)\psi_L +\bar\psi_L(\d_{\bz}-{1\over 2}
 {z \over 1+\bz z}+iV^*)\psi_R\}.\ee
 In the two-dimensional context, one often "gets rid" of the spin 
 connection by renormalizing the spinors $\bar\psi_D$, $\psi_D$:
 \be \psi_D=\sqrt{1+\bz z}\xi_D,\quad \bar\psi_D=\sqrt{1+\bz z}\bar\xi_D.\ee
 The action (124) then becomes 
 \be S_v={i\over 2\pi }\int d\bz \w dz \{\bar\xi_R(\d_z-iV)\xi_L +
 \bar\xi_L(\d_{\bz}+iV^*)\xi_R\}.\ee
 One should remember, however, that the spinors $\xi_R$,$\bar\xi_L$
 now belong to  a $\B$-bimodule $\B_1$ linearly 
 generated (over Grassmann numbers) by the elements of $\B$ of the form 
\be {\bz^{\bk}z^k\over (1+\bz z)^{m}},\quad max(\bar k, k+1)
\leq m,\quad  k,\bar k,m\geq 0\ee 
and $\xi_L$,$\bar\xi_R$ to a bimodule $\B_{\bar 1}$  generated by elements 
\be {\bz^{\bk}z^k\over (1+\bz z)^{m}},\quad max(\bar k+1, k)
\leq m,\quad  k,\bar k,m\geq 0.\ee 
The action of the massless spinor electrodynamics 
(the Schwinger model) on $S^2$ can be then written as
\be S={i\over 2\pi }\int d\bz \w dz\{\bar\xi_R(\d_z-iV) \xi_L +
\bar\xi_L(\d_{\bz}+iV^*)\xi_R
+{1\over 4g^2}(1+\bz z)^2(\d_{\bz}V-\d_z V^*)^2\}.\ee

Our next task will be to rewrite the spinor electrodynamics (129) in 
the form which would use the Hamiltonian vector fields $T_i,\bar T_i$
instead of $\d_z,\d_{\bz}$ 
and the (real) Hamiltonian 1-form $A\in\o_1$ instead of $v$. To do this we have 
first to encode the spinors $\bar\xi_D,\xi_D$ as elements of the algebra
$\A$ (cf.(1)). This is easy: define 
\be \Phi \equiv b\bar\xi_L+\bb\xi_L,
\quad \bar\Phi\equiv b\xi_R+\bb\bar\xi_R;\ee
the fact that  $\xi_R,\bar\xi_L\in\B_{1}$ and $\xi_L,\bar\xi_R\in\B_{\bar 1}$
implies that $\Phi$ and $\bar\Phi$ belong to $\A$. We also have to identify
how the 1-form $v$ enters into the real Hamiltonian 1-form $A$. This is given
as before in (99).
 With the ansatz (99) and (130), 
it turns out that the standard  Schwinger model on $S^2$
can be rewritten as 
\be S=-{1\over 4g^2}I[(dA)_0^2]+{1\over 2}I[\bT_2\bar\Phi\bT_1\Phi+
\T_2\bar\Phi\T_1\Phi +\bT_2\Phi\bT_1\bar\Phi+\T_2\Phi\T_1\bar\Phi]
 .\ee
 The quantity  $P$ (coming from (99)) 
 plays  the role of a nondynamical
  auxiliary field in (131); it can be eliminated by its equation of motion to
  yield (129). The meaning of the symbols in (131) is the following:
 $\bT_j,\T_j$ are covariant derivatives acting on $\Phi$ and  $\bar\Phi$
as follows
\be \T_j\Phi=T_j\Phi -iA_j\Phi,\quad \bT_j\Phi=\bar T_j\Phi-i\bar A_j\Phi,\ee
\be \T_j\bar\Phi=T_j\bar\Phi +\bar\Phi iA_j,\quad \bT_j\bar\Phi=
\bar T_j\bar\Phi+\bar\Phi i\bar A_j;\ee
the  integral $I$  was defined in (5).

It is easy to check that the action  (131)  has
 the gauge symmetry with the
gauge group $G_e$ (cf. (104)). An element $U$ of $G_e$ acts on 
$\Phi,\bar\Phi$ and $A$ as follows
\be \Phi\to U\Phi,\quad \bar\Phi\to \bar\Phi U^{-1},
\quad A\to A-idU U^{-1}.\ee

Now it is straightforward to write the action of the deformed Schwinger
model:
 \be S_N=-{1\over 4g^2}\ST[(dA-iA^2)_0^2]
+{1\over 2}\ST[\bT_2\bar\Phi\bT_1\Phi +\T_2\bar\Phi\T_1\Phi+
\bT_2\Phi\bT_1\bar\Phi+\T_2\Phi\T_1\bar\Phi].\ee
As in the case of
the scalar electrodynamics, here $A\in\o_{1N}$ is a real noncommutative
Hamiltonian 1-form and the deformed
spinor fields $\bar\Phi$ and $\Phi$ are elements of $\A_{aN}\oplus\A_{\bar aN}$
(cf. (66) and (67)) with Grassmann coefficients.
 The operators
$\T_j,\bT_j$ (by a little abuse of notation we denote them in the
same way as the undeformed quantities appearing in (131)) act as
\be  \T_j\Phi=N[t_j,\Phi]_+ -iA_j\Phi,\quad \bT_j\Phi=N[\bar t_j,\Phi]_+
-i\bar A_j\Phi;\ee
\be  \T_j\bar\Phi=N[t_j,\bar\Phi]_+ +\bar\Phi iA_j,\quad \bT_j\bar\Phi=
N[\bar t_j,\Phi]_+
+\bar\Phi i\bar A_j.\ee
The quantities $t_j,\bar t_j$ were defined in (9).

It is obvious that for $N\to\infty$ the deformed action (135) gives
the undeformed one (131). 
The gauge symmetry group  in the noncommutative case
is $G_{eN}$ (cf. (108)) and the deformed fields transforms as
\be \Phi\to U\Phi, \quad \bar\Phi\to \bar\Phi U^{-1},
\quad A\to UAU^{-1}-idU U^{-1}.\ee

A proof of the "Poincar\'e" symmetry $SU(2)$ of the undeformed action (131)
 and of the deformed one   (135) 
is easy: 1) the invariance of the  terms not containing
the fermions  was already proved
in the case of the scalar electrodynamics ; 2) the invariance of the
fermionic  terms follows from the $SU(2)$ invariance of the inner
products (5) and (69) and from the following commutation relations:
\be [R_3,T_1]={1\over 2}T_1, \quad [R_3,T_2]=-{1\over 2}T_2,
\quad [R_3,\bar T_1]=-{1\over 2}\bar T_1, \quad [R_3, \bar T_2]=
{1\over 2}\bar T_2;\ee
\be [R_+,T_1]=0,\quad [R_+,T_2]=T_1,\quad [R_+,\bar T_1]=-\bar T_2,\quad
[R_+,\bar T_2]=0;\ee
\be [R_-,T_1]=T_2,\quad [R_-,T_2]=0,\quad [R_-,\bar T_1]=0,\quad
[R_-,\bar T_2]=-\bar T_1.\ee
It is interesting to remark that  we can add to the Hamiltonian vector
fields $R_i, T_i,\bar T_1$ one more even vector field $Z$, generated by the 
Hamiltonian $\bar aa=({\bb b\over \bz z+1})$ and obeying
\be [Z,R_i]=0,\quad [Z,T_i]=-T_i,
\quad [Z,\bar T_i]=\bar T_i.\ee
Then the generators $R_i,T_i,\bar T_1$ and $Z$ fulfil the $osp(2,2)$
superalgebra commutation relations (47)-(50) and  (139)-(142).

Note that we can construct also the chiral electrodynamics by setting
the fields $\xi_L$ and $\bar\xi_R$  to zero.
In the noncommutative situation the latter case
 corresponds to saying that both
matrices $\Phi$ and $\bar\Phi$ are 
in $\A_{aN}$. The action will continue to be (131) in the commutative
 case and (135) in the noncommutative one and the gauge transformations
  will be (134) and (138), respectively. Thus we have achieved quite
  an interesting result: we have naturally coupled the gauge field
  to a chiral fermion while having only a finite number of degrees
  of freedom and no fermion doubling. Perhaps it would be somewhat premature
  to draw too optimistic conclusions from this two-dimensional story,
  nevertheless, there is a clear promise that the method might work
  also in higher dimensions.

\vskip2pc
\noindent{\bf Acknowledgement} 
\vskip1pc
\noindent I am grateful to A. Connes for enlightening discussions.

\end{document}